\input phys

\RF\bigi{I. Bigi, V. Khoze, N. Uraltsev, and A. Sanda, In {\it CP
Violation}, Ed. C. Jarlskog, World Scientific, Singapore.}
\centerline{\bf A Comment on Totally Inclusive $B^0$ Decays}
\centerline{L. Stodolsky}

\centerline{Max-Planck-Institute f\"ur Physik,
F\"ohringer Ring 6, 80802 Munich}

ABSTRACT:
We recall that the totally inclusive rate for $B^0$ decays can
yield information on CP violation.

\centerline{\bf -----------}
\bigskip

 One of the simplest measurements  in the $B^0$ system, which will
be under extensive study for evidence of CP violation   in the
coming years, is just the total, completely inclusive decay rate
as a function  of time.  

We would like to recall that such  a measurement yields information
concerning CP violation. Although the effect to be expected from
standard estimates is small, it may be worth experimental
consideration.

The point  we would like to recall is the following: if the totally
inclusive decay rate as a function of time of a $B^0$ (or a $K^0$
for that matter) deviates from the sum of two exponentials, then
CP is violated. This follows from the fact that such a deviation
is only possible if the two eigenstates of the mass matrix are not
orthogonal, which is not possible if CP is conserved.

That is, let
$$\psi=a_H(t)\vert H> +a_L(t)\vert L>$$
be the wavefunction representing the state of the $B$ system in
terms of states of definite mass and lifetime ``heavy" and
``light", so that $a_H$,$a_L$ are pure exponentials in time. The
square of $\psi$ represents the number of undecayed $B$'s and thus
gives the total number of decays up to a certain time. Now
with$\vert H>$ and $\vert L>$ normalized to one,
$$\vert \psi \vert ^2= \vert a_H\vert ^2+\vert a_L\vert ^2 +2
Re\{a_L^*a_H<L\vert H>\}.$$

 We thus see that deviations from the sum of two exponentials are
only possible if the two states are not orthogonal. On the other
hand if they are indeed non-orthogonal, then an oscillating factor
at the frequency of the mass difference is to be expected. Note
that these conclusions follow regardless of the initial values
$a_H(0)$,$a_L(0)$. This means that in a set-up where we  create the
$B^0$ state via a ``tag",  these statements are generally true with 
{\it any} tag, although the exact nature of the effect may vary
from case to case.
 
 A particularly simple case experimentally is the two-$B^0$ state
resulting from $\Upsilon(4S)\rightarrow BB$ with no tag at all;
that is we consider the total  rate, including {\it both} $B^0$'s.
At first it might be thought that there can be no effect, no
oscillatory terms at all, since we produce equal amounts of both
flavors and in the two dimensional $B^0$ space we begin with a
density matrix $\rho\sim I$. However, to follow the time
development we must write this density matrix in terms of the
possibly non-orthogonal states of definite lifetime:
$$\eqalign{
I=
{1\over 1-\vert<L\vert H>\vert^2}&\Big\{\vert H><H\vert~
+~\vert L><L\vert~\cr
&-~<H\vert L>~\vert H><L\vert~-~<L\vert H>~\vert
L><H\vert\Big\}\cr}$$

Thus  the total number of undecayed particles, given by the trace
of $\rho(t)$, contains in general an oscillatory term  ${1\over 1-
\vert<L\vert H>\vert^2}2 Re\{e^{-i(m_H-m_L^*)t} \vert<L\vert
H>\vert^2\}$. However, since $<L\vert H>$ is small, at least in
standard estimates, the effect here is the square of a small
quantity and so much less than that with a single or tagged beams
of $B$'s. Note that since  $B^+$, $B^-$ decays are exponential they
may even be left in the sample, if this is simpler experimentally.

 Non-orthogonality is not possible if CP is conserved,  for then
all states may be grouped into non-communicating CP even and odd
combinations, and the $B^0$ states of definite CP  remain
orthogonal. Hence an effect immediately implies CP violation.

 The converse, however is not true; CP violation may occur while
leaving the two states practically orthogonal. Indeed, this is the
situation anticipated by standard calculations. There the parameter
$\vert {q\over p}\vert-1$ giving the non-orthogonality is
estimated\quref{\bigi} to be $10^{-3}$ for $B_d$ and $10^{-4}$ for
$B_s$. Hence in the standard framework the  effect is predicted to
be observable with difficulty.

 On the other hand, the observation of a {\it substantial} effect
would be evidence not only of  CP violation but also of physics
beyond the  KM framework conventionally used to incorporate  CP
violation in the standard model.

I would like to thank V.Luth, A. Buras and H. Fritszch for a
helpful conversation and V. Zakharov for his assistance in
understanding these matters.
\refout
\bye